\documentclass[twocolumn,superscriptaddress,floatfix,aps,prl]{revtex4-1}
\usepackage{times}
\usepackage{gensymb}
\usepackage{amsmath}
\usepackage{graphicx}  % Include figure files
\usepackage{epstopdf}
\usepackage{dcolumn}  % Align table columns on decimal point
\usepackage{bm}  % bold math
\usepackage{color}
\usepackage{amssymb}   % for math
\usepackage{amsthm}
\usepackage{float}
\usepackage{multirow}
\usepackage{subfigure}
\usepackage{setspace}
\usepackage{pdfpages}
\usepackage{pgffor}

\makeatletter
\AtBeginDocument{\let\LS@rot\@undefined}
\makeatother

\newif\ifarXiv
\arXivtrue 
\begin{document}
%\date{\today}
\title{Dynamic nuclear spin polarization induced by Edelstein effect at Bi(111) surfaces}
\author{Zijian Jiang}
\author{V. Soghomonian}
\author{J. J. Heremans}
\email{heremans@vt.edu}
\altaffiliation{Author to whom correspondence should be addressed: Department of Physics, Virginia Tech, Blacksburg, VA 24061, USA}
\affiliation{Department of Physics, Virginia Tech, Blacksburg, Virginia 24061, USA}

\begin{abstract}

Nuclear spin polarization induced by hyperfine interaction and the Edelstein effect due to strong spin-orbit interaction is investigated by quantum transport in Bi(111) thin film samples. The Bi(111) films are deposited on mica by van der Waals epitaxial growth. The Bi(111) films show micrometer-sized triangular islands with 0.39 nm step height, corresponding to the Bi(111) bilayer height. At low temperatures a high current density is applied to generate a non-equilibrium carrier spin polarization by the Edelstein effect at the Bi(111) surface, which then induces dynamic nuclear polarization by hyperfine interaction. Comparative quantum magnetotransport antilocalization measurements indicate a suppression of antilocalization by the in-plane Overhauser field from the nuclear polarization and allow a quantification of the Overhauser field. Hence nuclear polarization was both achieved and quantified by a purely electronic transport-based approach.  

\end{abstract}
 
\maketitle

Spatial inversion symmetry exists in the Bi bulk but is broken normal to the surface, leading to strong Rashba-like spin-orbit interaction (SOI) due to the asymmetry of the surface-confinement potential for the two-dimensional (2D) surface states supported at the Bi(111) surface \cite{Koroteev2004,HofmannPrSfSci2006,Hirahara2007}. The Rashba parameter can reach $\approx$ 0.5 eV \AA, substantially larger than in e.g. InSb heterostructures \cite{Martin2011,RayInSb-PRB}. Bi thin films further show a high carrier mobility and a long mean free path \cite{Zhu2011}. The Bi (111) surface states have therefore been of recent interest \cite{FeldmanScience-2016,Du2016}. The Edelstein effect generates a non-equilibrium carrier spin polarization (CP) in materials with SOI in response to an applied electric field or a current density $\textbf{j}$, with the spin polarization direction normal to $\textbf{j}$ and the surface normal \cite{EdelsteinSSC1990,Borge2014,Shen2014,Pesin2012}. The Edelstein effect has its origin in spin-momentum locking due to SOI. The effect can be pronounced at surfaces and interfaces with strong SOI, such as the Ag/Bi(111) \cite{Rojas2013} and Cu/Bi(111) \cite{Isasa2016} interfaces. Given the strong SOI at the Bi(111) surface, an in-plane $\textbf{j}$ in a Bi thin film is expected to generate a non-equilibrium in-plane CP. In the present work the Edelstein effect appears as the most plausible dominant origin of the CP under application of $\textbf{j}$, rather than e.g. lateral or top-and-bottom spin Hall effects, as explained in Ref. \cite{SuppMatls}. Hyperfine interaction (HI) can by dynamic nuclear polarization (DNP) transfer the CP to a non-equilibrium in-plane nuclear spin polarization (NP). The present work shows such Edelstein-induced DNP, an example of the interplay between strong SOI, HI, and the Edelstein effect. The work also demonstrates that the effect of NP on quantum-coherent transport allows for a quantification of the polarization. The work is reminiscent of recent experiments where CP from the Edelstein effect generates a spin-transfer torque on magnetic moments \cite{EmoriPRB93-2016}, compared to this work where HI effectively mediates a spin-transfer torque on the nuclear spins. DNP from CP resulting from spin injection was previously predicted \cite{JohnsonAPL-2000} and the interplay between NP and CP from spin injection, mediated by HI, was studied in Fe/GaAs \cite{SalisPRB80-2009}. Another study used Faraday rotation to study DNP from current-induced NP in InGaAs \cite{TrowbridgePRB90-2014}. The present experiments however differ from the latter \cite{TrowbridgePRB90-2014} by using quantum magnetotransport measurements to quantify the DNP in an all-electrical setup, and by showing that the relatively higher carrier density in the Bi(111) surface states compared to semiconductors \cite{SalisPRB80-2009,TrowbridgePRB90-2014,PagetPRB15-1977,Optorientbook} allows DNP without application of an external magnetic field, relying only on the effective electronic field created by CP. 

HI refers to the coupling of carrier spins to the nuclear spins by an energy term $AI\cdot J$, where $A$ represents the hyperfine coupling constant \cite{Nisson2013,Tifrea2011}, $I$ the nuclear spin and $J$ the total carrier angular momentum. Two mechanisms contribute to HI \cite{Tifrea2011,Mukhopadhyay2015,Feher1959}, Fermi contact interaction (dominant when carrier and nuclear orbitals overlap \cite{Bucher2000}) and dipolar interaction \cite{Mukhopadhyay2015,Tifrea2011}. HI can be more pronounced for heavy atoms featuring atomic parameters with higher energy scales \cite{Nisson2013,Feher1959}, and for nuclei with large $I$. Both effects play a role strengthening HI for Bi, with $I$ = 9/2. Further, electrons in Bi have a substantial s-orbital component at the Fermi energy, $\sim$ 10\%, increasing the contact term and HI. The strong SOI in Bi may also enhance HI. Quantitative information on the strength of HI in semimetallic Bi is lacking. Yet experiments have studied the interaction between Bi donors in Si and the Si s-like conduction band carriers \cite{George2010,Morley2010,Feher1959}, concluding $A$ = 6.1 $\mu$eV. The Knight shift in Bi$_{2}$Se$_{3}$ shows $A$ = 27 $\mu$eV \cite{Nisson2013}. Such values for $A$ indicate that consequential HI is expected in semimetallic Bi as well as in Bi compounds. HI can lead to DNP where spin polarization is transferred from the carriers to the nuclei \cite{Economou2019,Maestro2013} and CP then generates NP. With NP established, the carriers experience HI as an effective in-plane magnetic field having the same effect as an external Zeeman field, the Overhauser field $B_{OH}$ \cite{Maestro2013,Tifrea2011,Tripathi2008,SuppMatls}. Similarly, via HI the electronic CP results in an in-plane effective magnetic field $B_e$ experienced by the nuclei \cite{Optorientbook,SuppMatls}. For DNP to occur, the dipole-dipole interaction field $B_L$ between neighboring nuclei ($B_L \approx$ 0.024 mT \cite{SuppMatls}) needs to be overcome by a nuclear Zeeman energy preventing a rapid T2 relaxation of NP \cite{SalisPRB80-2009,PagetPRB15-1977,Optorientbook,SuppMatls}. $B_L$ can be overcome by a sufficiently large $B_e$ \cite{Optorientbook}. In semiconductor experiments $B_e$ is low due to the low carrier density, and overcoming the decay of NP then requires an external magnetic field $> B_L$ \cite{SalisPRB80-2009,TrowbridgePRB90-2014,PagetPRB15-1977,Optorientbook}. In contrast, the present work shows that the higher carrier density in the Bi(111) surface states provides a $B_e \gg B_L$ so that DNP can occur without an in-plane external magnetic field, and in fact application of an in-plane field keeps results unchanged \cite{SuppMatls}. 

$B_{OH}$ and the NP are here quantified by the antilocalization (AL) quantum coherence corrections to the conductance of the Bi(111) surface states, caused by quantum interference between backscattered time-reversed carrier trajectories under SOI. At low temperatures $T$, the AL corrections lead to a resistance $R$ with a specific dependence on an external magnetic field $B_\perp$ normal to the surface \cite{Golub2005,DeoYaoTM,Bergmann2010}. The magnetoresistance (MR, $R(B_\perp)$) due to AL is determined by three characteristic times \cite{DeoYaoTM,Bergmann2010}: the elastic scattering time $\tau_{0}$ as deduced from the areal surface state density $N_S$ and mobility $\mu$, the SOI spin decoherence time $\tau_{SO}$, and the quantum phase decoherence time $\tau_{\phi}$. Here $\tau^{-1}_{SO} \propto \Delta_{SO}^2$ where $\Delta_{SO}$ denotes the SOI splitting at the Fermi wavevector. The times are experimentally determined by quantitative fitting of the MR data to the AL theory developed by Iordanskii, Lyanda-Geller and Pikus (ILP) \cite{Iordanskii1994} appropriate for the Bi(111) 2D surface states with Rashba-like SOI \cite{SuppMatls}. The influence of magnetization on AL in ferromagnetic materials has been theoretically studied \cite{Dugaev2001}. We expect similar effects due to NP, supported by the theoretical treatment of $B_{OH}$ as an effective in-plane magnetic field $B_\parallel$ \cite{Komnik2007,Malshukov1997}. Specifically, $B_\parallel$ generates an effective Zeeman splitting which aligns the carrier spins and hence suppresses the Cooperon in the spin singlet channel and thereby inhibits AL \cite{Dugaev2001}. The inhibition of AL is visible in the data as an increase in $\tau_{SO}$ with increasing $B_\parallel$. Further, AL is a sensitive probe of quantum and spin coherence \cite{DeoYaoTM}, and is sensitive to the time-reversal symmetry (TRS) breaking due to $B_\parallel$ \cite{Dugaev2001,Meijer2004,Altshuler1981}. The breaking of TRS due to the interplay of Zeeman splitting and SOI results in a quantifiable decrease in $\tau_{\phi}$ \cite{Meijer2004} with increasing $B_\parallel$, also visible in the data. Identifying $B_\parallel = B_{OH}$, we thus use AL as a sensitive probe of DNP and HI which allows a quantification of $B_{OH}$. 

\begin{figure}
	\centering
	\includegraphics[width=0.5\textwidth]{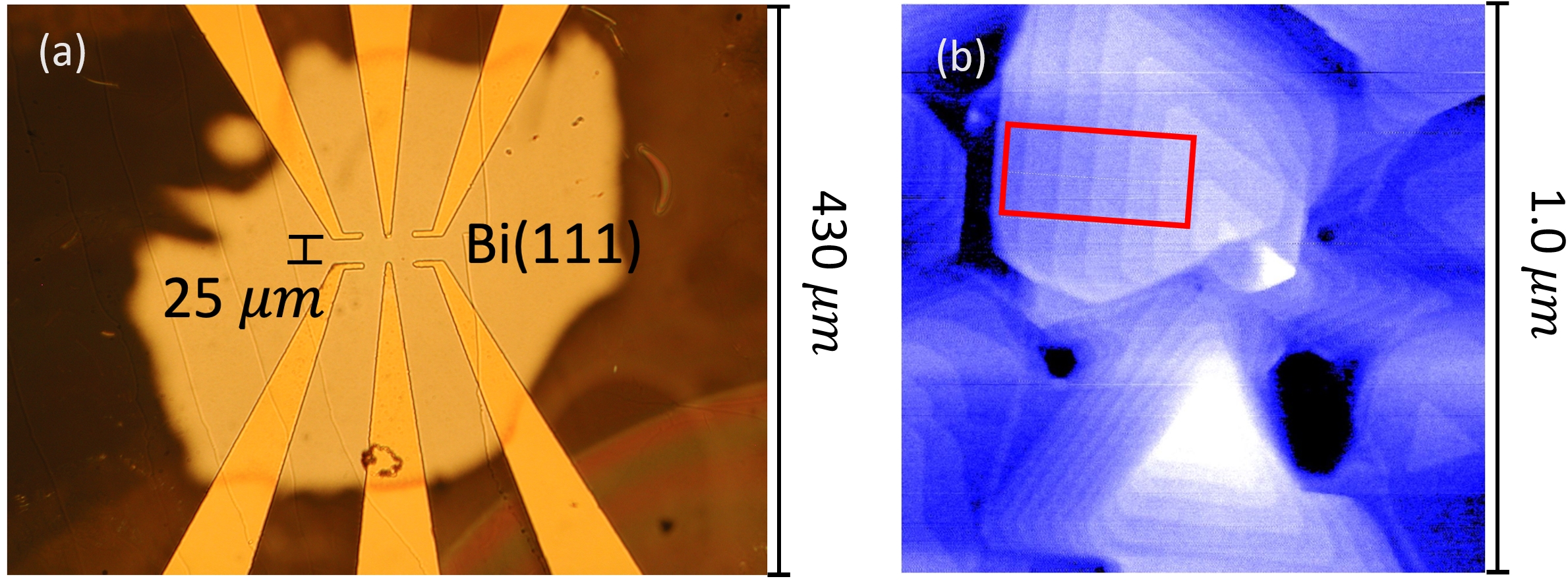}
	\caption{(a) Optical micrograph of the 40 nm thick Bi film sample grown on mica by van der Waals epitaxy, with lithographic Au contacts. The diameter of the sample is $\sim$ 350 $\mu$m; distance between contacts $\sim$ 25 $\mu$m. (b) AFM micrograph of a 1 $\mu$m $\times$ 1 $\mu$m region of the Bi film clearly illustrates layered growth. Step analysis in the red boxed region indicates a step height of 0.391 $\pm$ 0.015 nm, as expected for 1.0 BL$_{111}$.}
	\label{fig:AFM-SEMimage}
\end{figure}

An optimized van der Waals epitaxy (vdWE) \cite{Koma1999} was used to grow the Bi(111) films on mica substrates, resulting in large grain sizes with the trigonal axis perpendicular to the film plane \cite{SuppMatls}. vdWE is particularly suited to the unstrained growth of weakly bonding materials such as Bi \cite{Osten1991,Littlejohn2017}. The 40 nm thick Bi(111) was deposited through a shadowmask, yielding samples of diameter $\sim$ 350 $\mu$m. Au contacts were photolithographically patterned after film deposition (Fig.~\ref{fig:AFM-SEMimage}a). Atomic force microscopy indicated a layered step surface with triangular terraces (Fig.~\ref{fig:AFM-SEMimage}b) and showed a step height between adjacent terraces of 0.391 $\pm$ 0.015 nm, corresponding to one Bi(111) bilayer height (BL$_{111}$ = 0.39 nm) \cite{SuppMatls}.  

\begin{figure}
	\centering
	\includegraphics[width=0.5\textwidth]{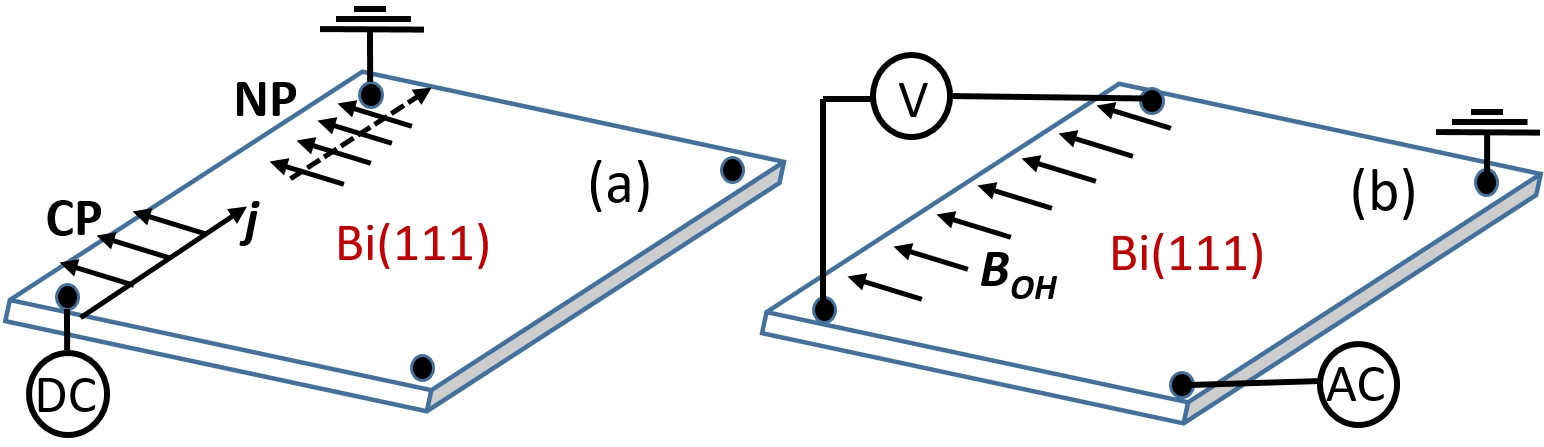}
	\caption{Schematic of the Edelstein-induced DNP and AL setup for Bi(111) surface states. (a) A high DC current density \textbf{j} in the Bi film sample induces a surface-state non-equilibrium carrier spin polarization by the Edelstein effect. The surface-state carrier spins are oriented perpendicular to \textbf{j}, and induce an in-plane surface nuclear spin polarization via DNP, resulting in in-plane Overhauser field $B_{OH}$. (b) After \textbf{j} is removed and while $B_{OH}$ slowly decays, AL measurements are carried out.}
	\label{fig:setup}
\end{figure}

The AL and transport coefficient characterization were carried out by magnetotransport in a $^3$He immersion cryostat down to $T$ = 0.39 K, using standard 4-contact AC lock-in techniques with current of 2 $\mu$A rms under applied $B_\perp$. To develop DNP a high DC polarization current, $I_p$ = 0.5 mA to 1.5 mA, \textbf{j} $\sim 6.25 \times 10^7$ A/m$^2$ to $1.9 \times 10^8$ A/m$^2$, was applied at $T$ = 0.39 K between a pair of contacts for variable polarization durations $t_p$ from 10 to 120 min. $I_p$ was removed after the DNP step, letting the NP and $B_{OH}$ decay slowly with a spin-lattice relaxation time T1 characteristic of the nuclear decoherence \cite{Yusa2005,Keane2011}. The slow decay allowed time for the subsequent observation of DNP from AL measurements. For AL measurements the voltage was measured over the same contacts to which $I_p$ was applied and hence over the path of which $B_{OH}$ develops, as depicted in Fig.~\ref{fig:setup}. For the AL data it is sufficient to sweep $B_\perp$ over $\sim$ 0.2 T, achievable in as little as $\sim$ 15 min, of the order of the expected T1 \cite{Lampel1968,Heil1995}. Experiments were also performed with different delay times $t_{delay}$, from 15 to 40 min, inserted between removing $I_p$ and performing the AL measurement, to characterize the decay in $B_{OH}$ and estimate T1. 

$N_S$ and $\mu$ were determined from magnetotransport at 0.39 K, indicating predominantly n-type surface carrier contribution. We determine $N_S$ = 1.95 $\times$ $10^{15}$ m$^{-2}$, $\mu$ = 1.00 m$^2$/Vs, $\tau_0$ = 0.0856 ps and mean free path $l_0$ = $v_f\tau_0$ = 20.4 nm, where $v_f$ is the Fermi velocity derived from $N_s$. As appropriate for surface states we use the 2D diffusion constant $D$ calculated as $D = \frac{1}{2} v_f^2\tau_0$, at $T$ = 0.39 K yielding $D$ = 0.00243 m$^2$/s. AL results in a characteristic positive quantum correction in $R(B_\perp)$ at $B_\perp \lesssim$ 0.4 T, expressed as a small correction to the 2D conductivity $\sigma_2(B)$. We define $\Delta\sigma_{2}(B_\perp)=\sigma_{2}(B_\perp)-\sigma_{2}(B_\perp=0)$ and $\Delta R(B_\perp) = R(B_\perp) - R_0$ where $R_0 = R(B_\perp=0)$. Since $\Delta R(B_\perp) \ll R_0$, we have $\Delta\sigma_{2}(B_\perp)/\sigma_{2}(B_\perp=0)\approx-\Delta R(B_\perp)/R_{0}$, allowing fits to $\Delta\sigma_{2}(B_\perp)$ from the experimental MR. To fit the data ILP theory \cite{Iordanskii1994} is applied, including only the Rashba SOI term (details in \cite{SuppMatls}). Since $\tau_{0}$ merely produces a shift in $\Delta\sigma_2(B_\perp)$, $\tau_\phi$ and $\tau_{SO}$ are the only two free fitting parameters. The fits are performed for AL obtained after different $t_p$ and $t_{delay}$ under different $I_p$. From the fits, we find the dependences on $t_p$, $t_{delay}$ and $I_p$ of $\tau_{SO}$ and $\tau_{\phi}$. From the latter the dependences of $B_{OH}$ are determined.  

\begin{figure}
	\centering
	\includegraphics[width=0.48\textwidth]{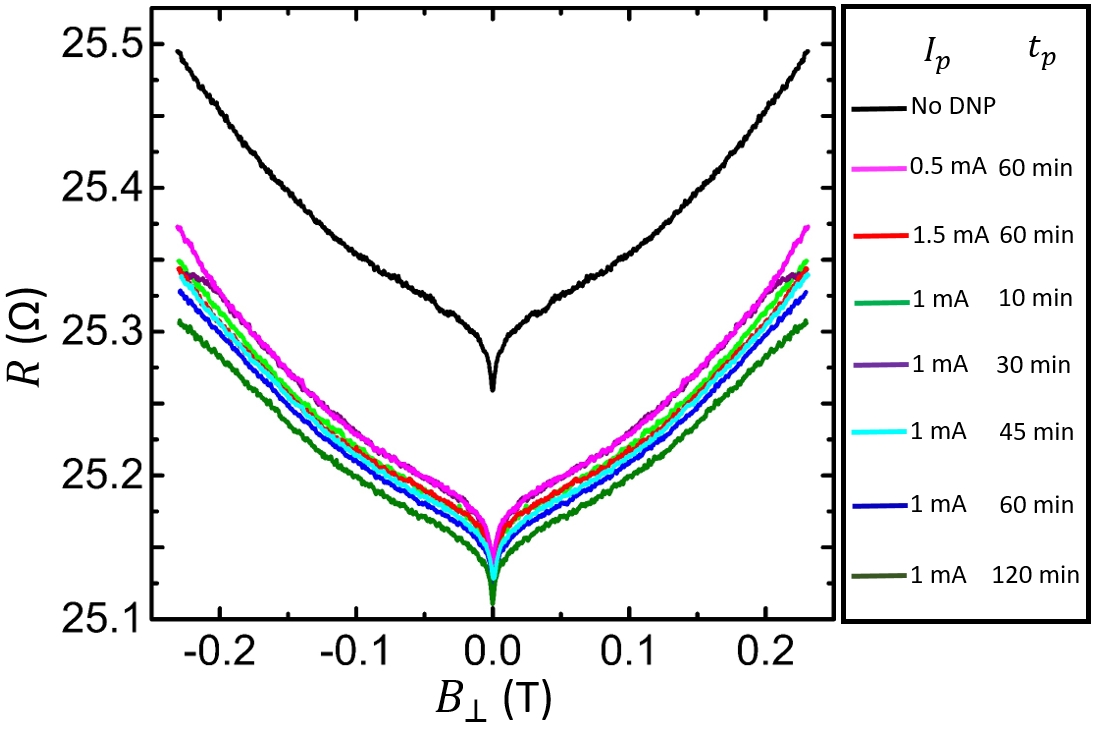}
	\caption{AL magnetoresistance at $T$ = 0.39 K before (indicated as No DNP) and after DNP with variable $I_p$ and variable $t_p$ ($t_{delay}=0$; traces not offset). After DNP a widening of $R(B_\perp)$ vs $B_\perp$ for $B_\perp \neq 0$ is evident.}
	\label{fig:Rxx}
\end{figure} 

\begin{figure}
	\centering
	\includegraphics[width=0.48\textwidth]{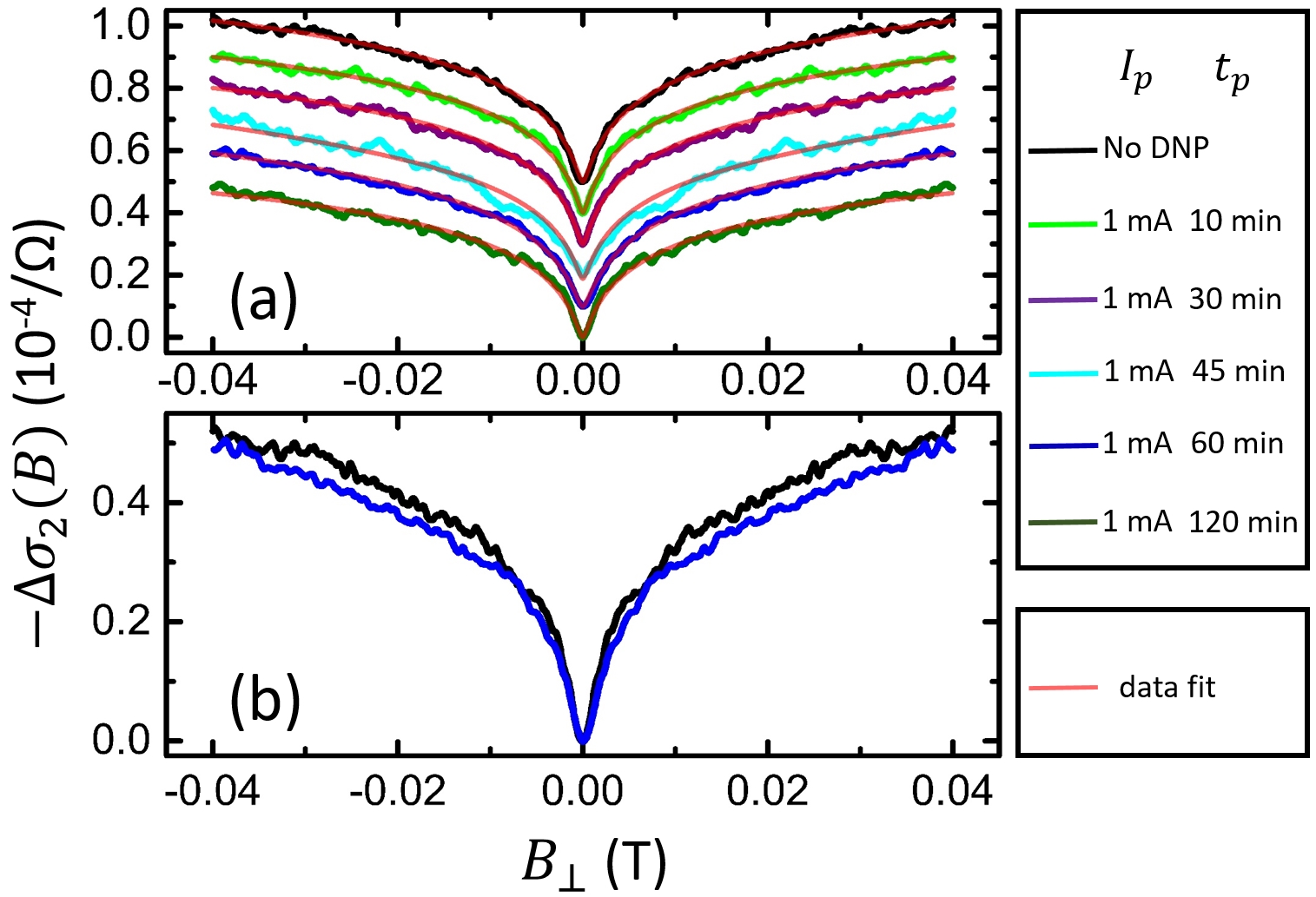}
	\caption{2D conductivity corrections due to AL at $T$ = 0.39 K and at low $B_\perp$ ($t_{delay}=0$): (a) under variable $t_p$ with $I_p$ = 1 mA. The red traces indicate fits to the AL theory \cite{Iordanskii1994}. Data are offset for clarity; (b) before DNP (black trace) and after DNP (blue trace) with $t_p$ = 60 min and $I_p$ = 1 mA (traces not offset). The widening of the trace after DNP indicates a partial suppression of AL by $B_{OH}$.}
	\label{fig:ILP}
\end{figure}

Figure~\ref{fig:Rxx} depicts representative MR of the Bi film sample at $T=0.39$ K before and after DNP using variable $I_p$ ranging from 0.5 mA to 1.5 mA and $t_p$ ranging from 0 (before DNP) to 120 min (at $t_{delay}=0$). The positive MR characteristic of AL is observed both before and after DNP. The negative of $\Delta\sigma_2(B_\perp)$ (reproducing $\Delta R(B_\perp)$) at low $B_\perp$ is displayed in Fig.~\ref{fig:ILP}a for variable $t_p$ when $I_p$ = 1 mA (at $t_{delay}=0$). Best fits to the ILP theory \cite{Iordanskii1994,SuppMatls} overlay the data in Fig.~\ref{fig:ILP}a in red and indicate that the theory excellently captures the AL in the Bi(111) surface states and will allow reliable extraction of values for $\tau_{SO}$ and $\tau_{\phi}$. The traces for $R(B_\perp)$ (Fig.~\ref{fig:Rxx}) and for $-\Delta\sigma_2(B_\perp)$ (Fig.~\ref{fig:ILP}a) show a widening vs $B_\perp$ for $B_\perp \neq 0$ after DNP, characteristic of an increase in $\tau_{SO}$ (decreasing effect of SOI) and a decrease in $\tau_{\phi}$ as confirmed below. The widening shows a dependence on $I_p$ and $t_p$, with long $t_p$ = 120 min at $I_p$ = 1 mA resulting in the largest effect. The dependence on $t_p$ and $I_p$ suggests DNP and hence $B_{OH}$ play a role in changing $\tau_{SO}$ and $\tau_{\phi}$. The widening of the minimum in $-\Delta\sigma_2(B_\perp)$ is further illustrated in Fig.~\ref{fig:ILP}b where the black trace represents $-\Delta\sigma_2(B_\perp)$ before DNP and the blue trace after DNP with $t_p$ = 60 min and $I_p$ = 1 mA (at $t_{delay}=0$). Before we present quantitative data on $\tau_{SO}$ and $\tau_{\phi}$, we note that the AL results after DNP are qualitatively consistent with the existence of in-plane $B_{OH}$. Phenomenologically, after removing $I_p$, $B_{OH}$ persists and generates an effective Zeeman energy $g_\parallel^*\mu_B B_{OH}$, where $g_\parallel^*$ denotes the in-plane $g$-factor (for Bi(111) surface states, $g_\parallel^* \approx 33$ \cite{Du2016}) and $\mu_B$ denotes the Bohr magneton. $B_{OH}$ partially aligns the carrier spins and suppresses the spin phase shift due to SOI and thereby weakens AL \cite{Dugaev2001,Meijer2004,MeijerPRL2005}. The effect leads to a widening of the characteristic sharp minimum in $\Delta R(B_\perp)$ vs $B_\perp$ and is quantified by a lengthening of $\tau_{SO}$. Further, $B_{OH}$ results in a spin-induced TRS breaking \cite{Malshukov1997,Meijer2004,MeijerPRL2005}, leading to a decrease in $\tau_{\phi}$. While it is not in the scope of this experimental study to modify the ILP theory to include HI, future theoretical studies specific to the influence of HI and NP on AL may help refine quantitative aspects of the experiments, as was performed for ferromagnetic order \cite{Dugaev2001} and for Zeeman interaction \cite{Malshukov1997}. 

The dependences of $\tau_{SO}$ and $\tau_\phi$ on $t_p$ at fixed $I_p$ = 1 mA with $t_{delay}=0$ are presented in Fig.~\ref{fig:tausotauphi}a-b. The value of $\tau_{SO}$ increases with increasing $t_p$ (Fig.~\ref{fig:tausotauphi}a), indicative of the influence of the in-plane $B_{OH}$. A phenomenological understanding was presented above. Theoretical studies of the combined influence of SOI and $B_\parallel$ on an inhomogeneous interfacial spin distribution \cite{Froltsov2001} show that even a weak $B_\parallel$ results in a decrease of the spin density proportional to 1/(2$\pi D \tau_{SO}$), relating an increase in $\tau_{SO}$ to the influence of $B_\parallel = B_{OH}$. Figure~\ref{fig:tausotauphi}b shows a decrease of $\tau_\phi$ with increasing $t_p$, and similar to Fig.~\ref{fig:tausotauphi}a manifests a saturation at higher $t_p$. The decrease of $\tau_\phi$ with increasing $t_p$ is indicative of the interplay of the effective Zeeman energy and SOI \cite{Malshukov1997,Meijer2004}, predicted to result in a quadratic dependence of $\tau_\phi$ on $B_\parallel$ \cite{Meijer2004}: 

\begin{figure}
	\centering
	\includegraphics[width=0.48\textwidth]{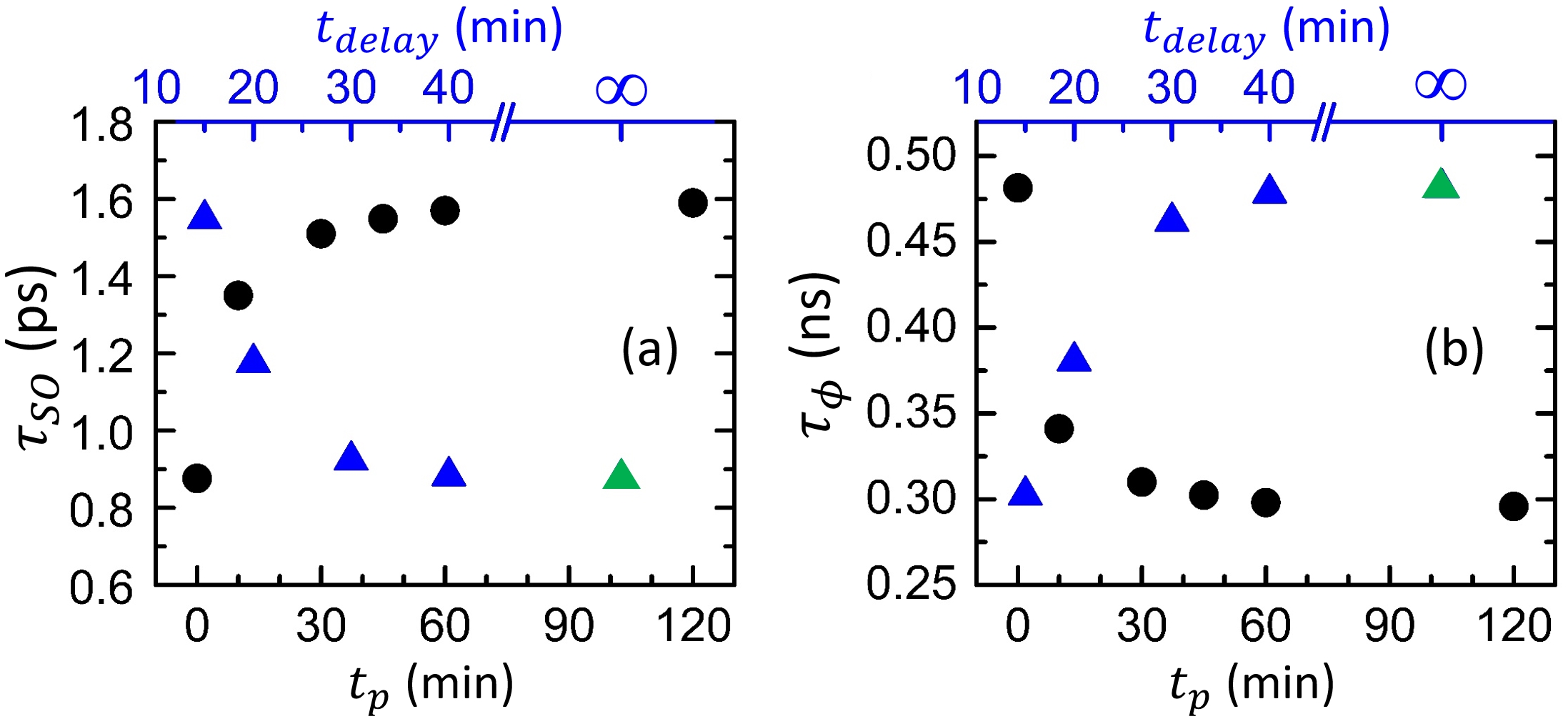}
	\caption{(a) Spin-orbit decoherence times $\tau_{SO}$ and (b) quantum phase decoherence times $\tau_\phi$ at $T$ = 0.39 K and $I_p$ = 1 mA, vs DNP duration $t_p$ ($t_{delay}=0$) (black circles) and vs $t_{delay}$ ($t_p$ = 60 min) (blue triangles). Data without DNP stand in for $t_p=0$ and for $t_{delay}\to\infty$ (green triangles).}
	\label{fig:tausotauphi}
\end{figure} 

\begin{equation}
\frac{\tau_\phi(B_\parallel)}{\tau_\phi(B_\parallel=0)} = \frac{1}{1 + cB_\parallel^2},
\label{eq:tauphi}
\end{equation}
where $c = \tau_\phi(B_\parallel=0)\tau_{SO}(B_\parallel=0)(g_\parallel^*\mu_B/\hbar)^2$. The estimated average value of $B_{OH}=B_\parallel$ can be calculated from the data using Eq.~\ref{eq:tauphi}. Figure~\ref{fig:tausotauphi}a-b depicts the dependences of $\tau_{SO}$ and $\tau_\phi$ on $t_{delay}$ at $I_p$ = 1 mA and $t_p$ = 60 min ($-\Delta\sigma_2(B_\perp)$ in \cite{SuppMatls}). With increasing $t_{delay}$, $\tau_{SO}$ decreases and $\tau_\phi$ increases to their values without DNP, consistent with a decay in $B_{OH}$. Figure~\ref{fig:BOH} shows the average $B_{OH}$ calculated from $\tau_\phi$ in Fig.~\ref{fig:tausotauphi}b. Since the AL measurement (sweeping over $B_\perp \sim$ 0.2 T after removing $I_p$ and waiting $t_{delay}$) spans $\sim$ 15 min, by estimated average $B_{OH}$ is meant the value after averaging over these $\sim$ 15 min. Current spreading between the current contacts over the sample geometry during DNP will likely lead to non-uniform DNP, and $B_{OH}$ hence encompasses spatial averaging as well. To minimize handling of the data, the averaging effects are not accounted for in Fig.~\ref{fig:BOH} but should be kept in mind. In Fig.~\ref{fig:BOH} the average $B_{OH}$ increases with increasing $t_p$, and saturates at about 13 mT. An exponential fit showed that the increase towards saturation occurs with a characteristic time T1e = 6 ... 11 min, with T1e characterizing the expected nuclear spin alignment by DNP \cite{Optorientbook}. In Fig.~\ref{fig:BOH}, the average $B_{OH}$ decays exponentially with increasing $t_{delay}$, with spin-lattice relaxation time T1 = 11.4 min. The value T1 = 11.4 min is of the order of expected values \cite{Lampel1968,Heil1995}. $B_{OH}$ depends on the average nuclear spin $I_{av}$ after NP, as $B_{OH}=AI_{av}/(g_\parallel^*\mu_B)$ \cite{PRB92Tenberg2015,Optorientbook,SuppMatls}, and $I_{av}$ follows a Brillouin function in the average carrier spin $S_{av}$ after CP \cite{Optorientbook,SuppMatls}. Using values of $A$ = 6.1 $\mu$eV to 27 $\mu$eV \cite{George2010,Morley2010,Feher1959,Nisson2013} we find that $B_{OH}=$ 13 mT is reached for $S_{av}= 0.37$ if $A$ = 6.1 $\mu$eV and for $S_{av}= 0.20$ if $A$ = 27 $\mu$eV \cite{SuppMatls}. Since we do not expect full NP ($S_{av}= \frac{1}{2}$) and $B_{OH}$ involves averages described above, the saturation value of 13 mT is consistent with the knowledge of $A$ in Bi and with plausible values of $S_{av}$. For $B_{OH}=$ 13 mT and in this range of $A$ it is calculated that $B_e \gg B_L$, consistent with the observation of DNP without external magnetic field \cite{SuppMatls}. Also, the dependence of $B_{OH}$ on $I_p$ strongly resembles the expected Brillouin function \cite{SuppMatls}, strengthening the consistency between expectations and data. The saturation value $B_{OH}=$ 13 mT and the dependences on $t_p$, $t_{delay}$ and $I_p$ firmly suggest that the CP due to the Edelstein effect was transferred by HI to the Bi nuclei, demonstrating Edelstein-induced DNP and its measurement by quantum transport. 

\begin{figure}
	\centering
	\includegraphics[width=0.37\textwidth]{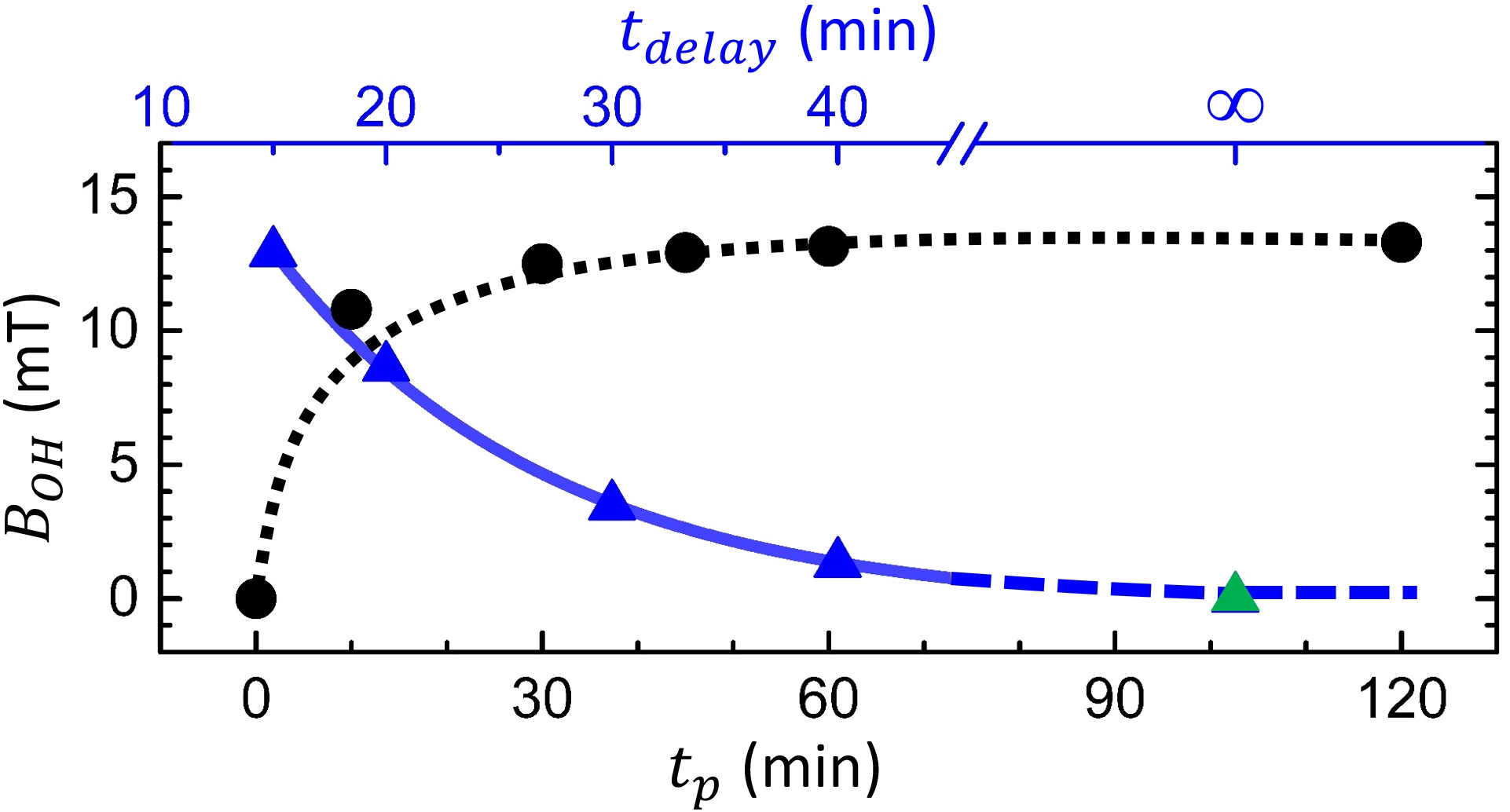}
	\caption{Overhauser field $B_{OH}$ at $T$ = 0.39 K and $I_p$ = 1 mA, vs DNP duration $t_p$ ($t_{delay}=0$) (black circles) and vs $t_{delay}$ ($t_p$ = 60 min) (blue triangles). Data without DNP stand in for $t_p=0$ and for $t_{delay}\to\infty$ (green triangle). The black dotted line is a guide to the eye. The blue line is an exponential fit yielding T1 = 11.4 min.}
	\label{fig:BOH}
\end{figure} 

In conclusion, Bi(111) thin films were deposited by van der Waals epitaxy on mica substrates. Using antilocalization quantum-coherent transport measurements on the Bi(111) surface states to detect in-plane magnetic fields, quantitative evidence was obtained for a transfer of carrier spin polarization to Bi nuclear spin polarization by hyperfine interaction. The carrier spin polarization was obtained via the Edelstein effect in the Bi(111) surface states. The experiments verify the existence of Edelstein-induced dynamic nuclear polarization, in an example of interaction between spin-orbit interaction and hyperfine interaction via the nuclear spin bath, with possible applications in nuclear spintronics and to polarize nuclei to mitigate spin decoherence via HI in quantum devices. The experiments also show that antilocalization forms a sensitive probe for hyperfine interaction and nuclear polarization. 

\section{\label{sec:acknowledgments}Acknowledgments}

The work was supported by the U.S. Department of Energy, Office of Basic Energy Sciences, Division of Materials Sciences and Engineering under award DOE DE-FG02-08ER46532.

\ifarXiv


\section*{Supplementary material}

# Supplemental Material

**Dynamic nuclear spin polarization induced by Edelstein effect at Bi(111) surfaces**

Zijian Jiang, V. Soghomonian and J. J. Heremans
Department of Physics, Virginia Tech, Blacksburg, Virginia 24061, USA

## S1: Van der Waals shadowmask epitaxy of Bi(111) thin films

High quality Bi thin film growth is challenging [1]. In the present work an optimized van der Waals epitaxy (vdWE) [2-4] is used, adaptable to growth of various 2D materials. Unlike conventional Stranski-Krastanov epitaxial growth where the bonding or interaction between the substrate and epilayer is often covalent or ionic, in vdWE the interaction is non-bonding and hence weak. vdWE is a choice when the substrate and/or the epilayer possess a van der Waals surface without dangling bonds, realized in 2D materials with naturally completely terminated surfaces, such as graphene and mica [2-4]. Epilayers of Sb, Ge and Ge/Sb on mica exhibit high crystalline quality [3,4], leading to mica as the substrate for the present Bi thin film growth. As the interaction between the deposited Bi layer and mica is weak, when the epitaxial Bi layer is deposited on the lattice-mismatched mica substrate (monoclinic, $a_{mica}$ = 519 pm, $b_{mica}$ = 904 pm), it at the outset grows unstrained, with a lattice constant of $a_{Bi}$ = 454 pm, the bulk lattice constant of Bi in a plane normal to the trigonal axis. The Bi films in this work grow by vdWE from the coalescence of isolated three-dimensional triangular islands, where each island grows layer-by-layer with a step height of 0.39 nm corresponding to one Bi(111) bilayer height ($BL_{111}$ = 0.39 nm), as detailed below.

The Bi films grow on mica with the trigonal axis (*c*-axis) perpendicular to the mica *ab* surface, yielding a Bi(111) surface. Onto a freshly cleaved mica substrate, high-purity (99.999 %) Bi was thermally evaporated at a base pressure of $10^{-8}$ Torr at room temperature. To remove absorbed surface water, the mica substrate was preheated at 250 °C for at least 24 hr under ultra-high vacuum before Bi deposition [2]. An Al shadowmask with apertures of diameters ~ 350 μm was placed on the mica surface. A 40 nm thick layer of Bi (~ 100 $BL_{111}$) was deposited through the apertures at a rate of 0.35 $BL_{111}$ /min [5]. The deposition rate and film thickness were monitored by a quartz microbalance monitor to within an accuracy of < 5 %. After deposition, the film samples were annealed at 95 ± 5 °C for 1 hr and then left in the chamber to cool before venting with dry nitrogen. On the resulting shadowmasked Bi film samples of diameter ~ 350 μm, photolithographically patterned Au was applied as contacts with a representative distance of 25 μm between two contacts (main text Fig. 1a).

As depicted in Fig. S1.1, scanning electron microscopy (SEM) of a representative area indicates large-scale uniformity and large grain size up to ~ 1 μm, with few defects. The SEM micrograph also shows characteristic triangular or hexagonal growth patterns (highlighted in red contours), corroborating the rhombohedral crystal structure of Bi thin films with the trigonal axis (*c*-axis) perpendicular to the mica surface. The triangular growth patterns are also apparent in the AFM micrograph in main text Fig. 1b, repeated here (Fig. S1.2b).

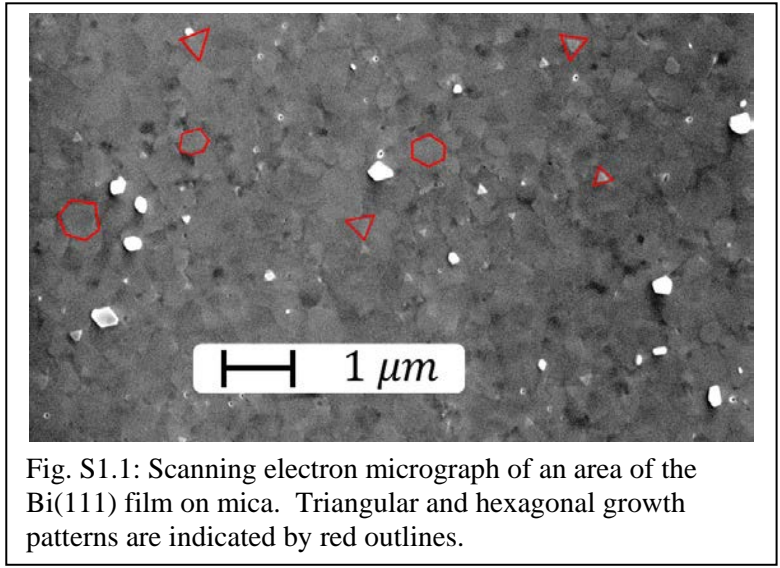

Fig. S1.1: Scanning electron micrograph of an area of the Bi(111) film on mica. Triangular and hexagonal growth patterns are indicated by red outlines.

Atomic force microscopy (AFM) clearly indicated a layered step surface with triangular terraces (main text Fig. 1b, Fig. S1.2b). As shown in Fig. S1.2a, the step height between adjacent terraces is measured to be $0.391 \pm 0.015$ nm, corresponding to 1.0 $BL_{111}$. Figure S1.2c depicts the Bi structure, highlighting the Bi(111) bilayers, the trigonal axis $c$, and a unit cell. Over a 1 μm x 1 μm, AFM measurements reveal that the root mean square roughness of the film was 1.53 nm. Compared to the sample thickness of 40 nm, the roughness measurement implies that the Bi film sample features a flat high-quality surface.

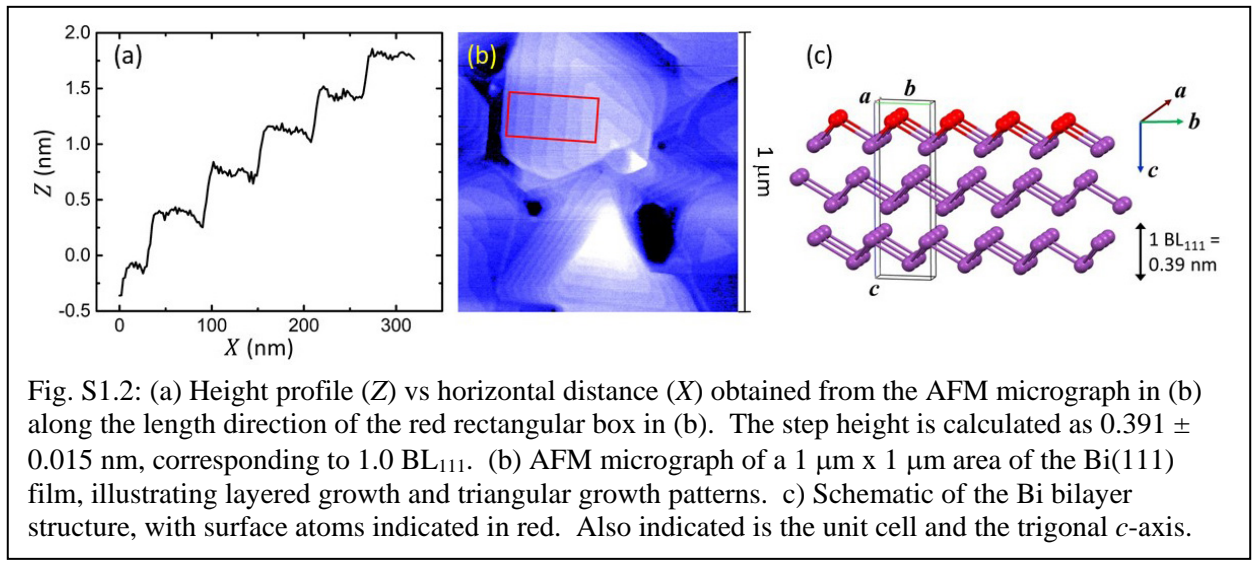

Fig. S1.2: (a) Height profile ($Z$) vs horizontal distance ($X$) obtained from the AFM micrograph in (b) along the length direction of the red rectangular box in (b). The step height is calculated as $0.391 \pm 0.015$ nm, corresponding to 1.0 $BL_{111}$. (b) AFM micrograph of a 1 μm x 1 μm area of the Bi(111) film, illustrating layered growth and triangular growth patterns. c) Schematic of the Bi bilayer structure, with surface atoms indicated in red. Also indicated is the unit cell and the trigonal $c$-axis.

## S2: Role of the Edelstein effect in charge-current to spin conversion

Because ferromagnetic materials are absent in the experiments, and because Bi interfaces are known to show strong Rashba-like spin-orbit interaction (SOI) [6,7] and the Edelstein effect [8,9], the latter effect stands out as the origin of the carrier spin polarization. The following points reinforce that view.

SOI is absent in bulk Bi (between the top and bottom interfaces) due to the existence of inversion symmetry in the bulk. Hence a spin Hall effect developed in the bulk Bi between top and bottom interfaces would be difficult to explain, and the origin of the carrier spin polarization must be sought at the interfaces.

A spin Hall polarization at top and bottom interfaces of the Bi(111) film is less likely as the origin. The top and bottom interfaces would, due to the spin Hall effect, have opposite spin accumulations, which will partially cancel their effect on nuclear spin polarization. The cancellation would likely not be complete, since the interfaces are different: the top interface is likely between Bi and a Bi oxide, the bottom interface a van der Waals interface with the mica substrate. Yet, the cancellation is still expected to lead to a much diminished effect if a spin Hall carrier polarization stood at the origin of the effect.

A lateral spin Hall polarization at the edges of the device due to SOI at the Bi top and bottom interfaces is in principle possible. But, the geometry does not have well-defined edges as it is not a lithographically prepared Bi mesa but a Bi flake with contacts on top. And, the antilocalization (AL) measurements sample not only the carrier population at the device edges but average the signal over the entire plane within which the current spreads. Hence in case of lateral spin Hall polarization the AL measurement would only return a diluted signal.

That leaves the Edelstein effect as the most plausible origin for the main part of the dynamic nuclear polarization signature. Both top and bottom Bi interfaces are expected to show an Edelstein effect. While our experiments cannot differentiate between dynamic nuclear polarization signatures from top and bottom interfaces, the top interface of the Bi film is expected to be less disordered, and hence contribute more to transport signatures such as the present dynamic nuclear polarization signature.

## S3: Analysis of antilocalization data

We performed quantitative analysis of the AL data using the theory of Iordanskii, Lyanda-Geller, and Pikus (ILP) [10], Eq. 13. This theory was used because it takes into account Rashba-like SOI due to spatial symmetry breaking normal to the surface containing the two-dimensional surface states [11] (SOI is absent in bulk Bi due to inversion symmetry). In Ref. [10], we set $\Omega_3 = 0$ (no cubic Rashba and Dresselhaus terms), and obtain the quantum correction to the 2D conductivity:

$$\sigma_2(B_\perp) = -\frac{e^2}{4\pi^2\hbar}\left\{\frac{1}{a_0} + \frac{2a_0+1+\frac{H_{SO}}{B_\perp}}{a_1\left(a_0+\frac{H_{SO}}{B_\perp}\right)-\frac{2H_{SO}}{B_\perp}} - \sum_{n=0}^{\infty}\left(\frac{3}{n} - \frac{3a_n^2+\frac{2a_n H_{SO}}{B_\perp}-1-\frac{2(2n+1)H_{SO}}{B_\perp}}{\left(a_n+\frac{H_{SO}}{B_\perp}\right)a_{n-1}a_{n+1}-\frac{2H_{SO}}{B_\perp}[(2n+1)a_n-1]}\right) + 2\ln\frac{H_0}{B_\perp} + \Psi\left(\frac{1}{2}+\frac{H_\phi}{B_\perp}\right) + 3C\right\},$$ where $a_n = n + \frac{1}{2} + \frac{H_\phi}{B_\perp} + \frac{H_{SO}}{B_\perp}$, $H_\alpha = \frac{\hbar}{4eD\tau_\alpha}$ with $\alpha = 0$, or $\phi$, or $S$.

Here $\tau_0$ denotes the elastic scattering time, $\tau_\phi$ the quantum phase decoherence time, $\tau_{SO}$ the SOI spin decoherence time, and the $H_\alpha$ denote characteristic magnetic fields, while $\Psi$ is the digamma function. The 2D diffusion constant $D = 0.00243$ m$^2$/s and $\tau_0 = 0.0856$ ps are obtained from conventional longitudinal and Hall resistance measurements (as explained in the main text; values at $T = 0.39$ K). In two dimensions as appropriate for surface states we have $D = \frac{1}{2} v_f^2 \tau_0$, where $v_f$ denotes the Fermi velocity. Therefore $\tau_\phi$ and $\tau_{SO}$ are the only two remaining free fitting parameters.

To approximate the sum $\sum_{n=0}^{\infty} \left( \frac{3}{n} - \frac{3a_n^2 + \frac{2a_n H_{SO}}{B_\perp} - 1 - \frac{2(2n+1)H_{SO}}{B_\perp}}{\left(a_n + \frac{H_{SO}}{B_\perp}\right)a_{n-1}a_{n+1} - \frac{2H_{SO}}{B_\perp}[(2n+1)a_n - 1]} \right)$, we set the upper limit of $n$ to 30000, which is sufficient as tests showed.

The ILP approach is suitable for the regime where $\Omega \tau_0 < 1$, with $\Omega$ representing the spin precession frequency due to SOI. The SOI energy splitting at the Fermi wavevector is expressed as $\Delta_{SO} = \hbar \Omega$ (note that in [10], $\Omega_1 = \frac{1}{2} \Omega$). An approach by Golub *et al.* [12] is in theory more appropriate than the ILP approach if $\Omega \tau_0 > 1$. The Golub approach was developed as a refinement of ILP for cases where SOI is strong (high $\Omega$) or mobility is high (long $\tau_0$, e.g. for ballistic transport or as appropriate for III-V semiconductors). We obtain values for $\Omega$ from the expression $1/\tau_{SO} = \frac{1}{2} \Omega^2 \tau_0$ [11] using $\tau_{SO}$ from the main text. With $\tau_0 = 0.0856$ ps, taking $\tau_{SO} \approx 1$ ps (main text Fig. 5(a)), we find $\Omega \tau_0 = (2\tau_0/\tau_{SO})^{1/2} \approx 0.4 < 1$. Hence the use of the ILP approach is justified. While SOI in the present Bi surface states is strong (high $\Omega$), the carrier mobility is lower than in e.g. III-V semiconductors (shorter $\tau_0$), rendering ILP a satisfactory formalism.

The ILP analysis should be restricted to magnetic fields $B_\perp$ below the characteristic field $H_0 = \hbar / (4eD\tau_0) = \hbar / (2e\, l_0^2)$, where $l_0 = v_f \tau_0$ denotes the mean free path. With $l_0 = 20.4$ nm, we find $H_0 = 0.79$ T, well above the range $-0.04$ T $< B_\perp < 0.04$ T we use for the analysis. The magnetoresistance (MR) data was obtained in each case with $B_\perp$ ranging over $\pm 0.2$ T, while the ILP fitting was performed only over the subrange $\pm 0.04$ T. The wider experimental range $\pm 0.2$ T was used out of caution, to ascertain that the sample's behavior had not changed in unaccountable ways that would indicate a lack of continuity in the data series, even though this did slow down the measurements. The fitting range was restricted to $\pm 0.04$ T to make sure no other MR phenomena, such as the almost inevitable geometrical MR, would contaminate the analysis. Yet for the ILP fitting it is important to capture the characteristic sharp dip in $R(B_\perp)$ at $B_\perp \approx 0$, as well as the gradual lessening of $dR/dB_\perp$ at higher $B_\perp$. The range $\pm 0.04$ T proved optimal to avoid other MR phenomena as well as to capture AL features necessary for the ILP fit, and amply satisfies the criterion $B_\perp < H_0$.

## S4: Effective hyperfine fields

The nonequilibrium nuclear spin polarization (NP) in this work, and hence the effective nuclear Overhauser magnetic field $\boldsymbol{B_{OH}}$ experienced by the electrons, result from the generation of a nonequilibrium electron spin polarization. Yet for dynamic nuclear spin polarization (DNP) to occur, the dipole-dipole interaction field $B_L$ between neighboring nuclei needs to be overcome ($B_L$ is typically a fraction of mT). Unless overcome by a nuclear Zeeman energy, this interaction will lead to a rapid relaxation of the nonequilibrium nuclear spin, with a relaxation time T2 ~ 0.1 ms [13-15]. The characteristic time for development of the NP by hyperfine interaction with

electrons is denoted T1e. Since T1e >> T2, the NP can be ignored unless $B_L$ responsible for the T2 relaxation is overcome by an actual or effective magnetic field $B_{eff}$ experienced by the nuclei, requiring $B_{eff} >> B_L$ [13-15]. Further, given $B_{eff}$, the nuclear spin system is effectively isolated from the lattice because the nuclear spin-lattice relaxation is characterized by a time T1 >> T2 (the isolation of the nuclear spin system from the lattice allows the definition of a nuclear spin temperature, as distinct from lattice temperature or electron temperature) [15]. The average nuclear spin after polarization is given by $I_{av}$:

$I_{av} = I\, B_I(x)$,  $x = I \ln[((1+2S_{av})/(1-2S_{av}))\,((1+2S_{th})/(1-2S_{th}))]$,
$S_{th} = (1/2) \tanh[(\mu_B\, g_\| B)/(2 k_B T)]$

where $I = 9/2$ is the Bi nuclear spin, and $B_I(x)$ is the Brillouin function for $I$ [15]. $S_{av}$ is the average electron spin after electron spin polarization, $S_{th}$ is the equilibrium value of the average electron spin at $B = B_{OH}$ and temperature $T$, $\mu_B$ is the Bohr magneton and $g_\|$ the in-plane g-factor ($g_\| \approx 33$ for Bi(111) surface states [5]). $S_{av}$ is limited by $S_{th} < S_{av} < ½$. At the value $B_{OH} = 13$ mT obtained from the experiments and at $T = 0.39$ K, we find $S_{th} = 0.177$. $\boldsymbol{I_{av}}$ is colinear with $\boldsymbol{S_{av}}$. We note that for the limits $S_{av} \to S_{th}$ we have $I_{av} \to 0$, and for $S_{av} \to ½$ we have $I_{av} \to I = 9/2$.

An estimate of $B_L$ can be obtained [15] from the dipole expression $B_L = (\mu_0/4\pi)(\mu_I/a^3)$, where $\mu_I$ denotes the nuclear magnetic moment with $\mu_I = I\, \mu_N$ (with $\mu_N$ the nuclear magneton and $I = 9/2$), $a$ denotes the interatomic distance in Bi and $\mu_0$ the permeability of vacuum. Using $a \approx a_{Bi} = 454$ pm, the bulk lattice constant of Bi in a plane normal to the trigonal axis, we find $B_L \approx 0.024$ mT.

The hyperfine interaction is experienced by the electrons as an in-plane $B_{OH}$ yielding an effective Zeeman energy, described by $g_{//}\, \mu_B\, B_{OH} = A\, I_{av}$, where $A$ denotes the hyperfine coupling constant [16,17]. $\boldsymbol{B_{OH}}$ is colinear with $\boldsymbol{I_{av}}$. In case of Fermi contact interaction, $A$ can be expressed as [15]:

$A = (4/3)\, \mu_N\, \mu_B\, \mu_0\, \eta\, N$,

where $\eta$ denotes the squared Bloch wave function amplitude at the site of the nucleus and $N$ the volume density of nuclei in the material. In Bi, experiments have concluded $A \approx 6.1\ \mu eV\ ...\ 27\ \mu eV$ [16,18-20]. We note that $\eta$ can be large ($\eta \approx 10^3 ... 2\times10^4$) because the electron density has a sharp maximum at the nucleus. In Bi, $N = 2.82 \times 10^{27}$ nuclei/m$^3$. For $A \approx 6.1\ \mu eV\ ...\ 27\ \mu eV$ we find $\eta \approx 4.41\times10^3\ ...\ 1.95\times10^4$, within the range of expectations. We rewrite:

$B_{OH} = A\, I_{av}/(g_{//}\, \mu_B) = (4/3)\, \mu_N\, \mu_0\, \eta\, N\, I_{av}/g_\|$ .

Similarly, spin-polarized electrons result via hyperfine interaction in an in-plane magnetic field $B_e$ experienced by the nuclei, expressed as [15]:

$B_e = -(4/3)\, \mu_B\, \mu_0\, \eta\, n_e\, S_{av}$,

where $n_e$ denotes the electron density (here estimated for the Bi(111) surface states). We hence have:

$B_e = -(A\, S_{av}/\mu_N)(n_e/N)$,

which can be compared to $B_{OH} = A\, I_{av}/(g_{//}\, \mu_B)$. $\boldsymbol{B_e}$ is colinear with $\boldsymbol{S_{av}}$. Hence, $\boldsymbol{S_{av}}$, $\boldsymbol{I_{av}}$, $\boldsymbol{B_{OH}}$ and $\boldsymbol{B_e}$ are all colinear. It was found experimentally that if the field $B_e$ generated by spin-polarized electrons is sufficiently large, then $B_e$ functions as a $B_{eff}$ surmounting dephasing by $B_L$, and

allows for DNP [15]. In experiments on semiconductors, where the carrier density is low, typically we have $n_e/N \ll 1$, and the effect of $B_e$ is negligible ($B_e < B_L$). In semiconductors, application of a small external magnetic field is hence necessary to obtain NP [13-15,21]. However, in semimetals such as Bi, carrier densities are substantially higher and $n_e/N$ is larger. As we will see, this results in $B_e \gg B_L$, allowing DNP to occur. The areal electron surface state density $N_S = 1.95 \times 10^{15}$ m$^{-2}$, as determined from magnetotransport at $T = 0.39$ K. Assuming the surface states are localized in the top Bi bilayer (an approximation) of thickness 0.39 nm, we obtain $n_e \approx 5.00 \times 10^{24}$ m$^{-3}$. This yields $n_e/N \approx 1.8 \times 10^{-3}$.

Values for $I_{av}$, $B_{OH}$ and $B_e$ depend on values for $A$ and/or $S_{av}$. $A$ is only known within a range from the literature, while $S_{av}$ is in the experiment not independently determined. The interaction between the spin-polarized electron current and the nuclei is also subject to nonuniformity due to current spreading. Yet, we show below that the ranges 6.1 µeV $< A <$ 27 µeV and $S_{th} < S_{av} <$ ½ yield $B_{OH}$ values compatible with $B_{OH} \approx 13$ mT obtained in the experiments, and that $B_e \gg B_L \approx$ 0.024 mT in all cases. Figure S4.1 shows $B_{OH}$ plotted vs $S_{av}$ ($S_{th} = 0.177 < S_{av} < 0.40$) for $A =$ 6.1 µeV and $A = 27$ µeV, with $S_{th} = 0.177$, where the level $B_{OH} = 13$ mT is indicated. Figure S4.2 shows $B_e$ plotted vs $S_{av}$ ($S_{th} = 0.177 < S_{av} < 0.4$) for $A = 6.1$ µeV and $A = 27$ µeV.

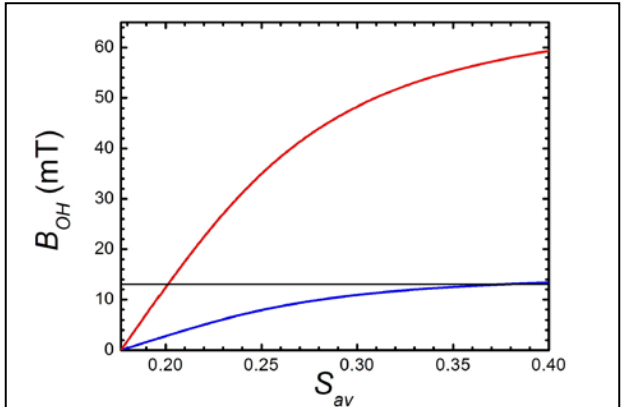 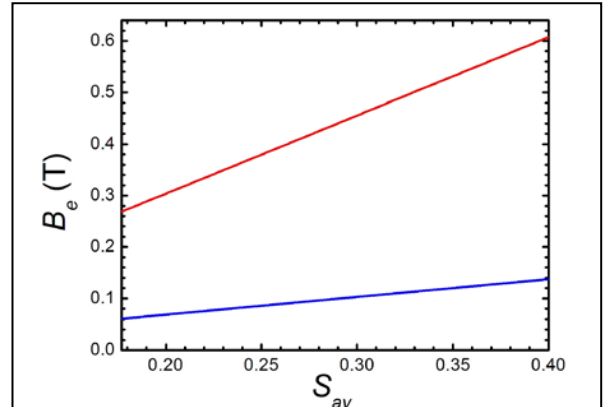

Fig. S4.1: Calculated $B_{OH}$ vs $S_{av}$ for $A = 6.1$ µeV (blue) and $A = 27$ µeV (red), with $B_{OH} = 13$ mT indicated as a black line. At $A = 6.1$ µeV, $B_{OH} =$ 13 mT is reached at $S_{av} = 0.37$. At $A = 27$ µeV, $B_{OH} = 13$ mT is reached at $S_{av} = 0.20$.

Fig. S4.2: Calculated $B_e$ vs $S_{av}$ for $A = 6.1$ µeV (blue) and $A = 27$ µeV (red). At $A = 6.1$ µeV, $S_{av} = 0.37$ yields $B_e = 0.129$ T. At $A = 27$ µeV, $S_{av} = 0.20$, yields $B_e = 0.305$ T. In both cases, $B_e \gg B_L$.

Figures S4.1 and S4.2 show that at $A = 6.1$ µeV, we have $B_{OH} = 13$ mT for $S_{av} = 0.37$, yielding $B_e = 0.129$ T. At $A = 27$ µeV, we have $B_{OH} = 13$ mT for $S_{av} = 0.20$, yielding $B_e = 0.305$ T. Both values for $S_{av}$ are realistic, and in both cases $B_e \gg B_L \approx 0.024$ mT. In Fig. S4.3 we have plotted $B_e$ vs $A$ assuming $S_{th} = 0.177$ and $B_{OH} = 13$ mT, showing that all values of $A$ result in $B_e \gg B_L$.

The calculations hence show that in the experiments the field $B_e$ generated by spin-polarized electrons is amply sufficiently large to surmount dephasing by $B_L$, and to allow for DNP. The observed $B_{OH} = 13$ mT is also consistent with the present knowledge of $A$ in Bi.

As a test, we performed DNP measurements with an external in-plane magnetic field $B_∥$ applied during the nuclear polarization, with $B_∥ = 0.1$ T and 1.0 T ($B_∥ ≈ B_e$ while $B_∥ \gg B_L$,). $\mathbf{B_∥}$ was applied in-plane and normal to the average current density direction of $I_p$, hence colinear with the expected $S_{av}$, $\mathbf{I_{av}}$, $\mathbf{B_{OH}}$ and $\mathbf{B_e}$. The in-plane field measurements were performed at $T = 1.30$ K, the lowest $T$ in the system allowing in-plane fields, and are depicted in Fig. S4.4. The figure contains the AL MR (negative of conductivity correction $\Delta\sigma_2$) plotted as $-\Delta\sigma_2(B_⊥)$ vs $B_⊥$, where $B_⊥$ denotes the magnetic applied normally to the surface. Figure S4.4 contains a comparison MR trace where no DNP was performed, three MR traces with DNP under $I_p = 1$ mA applied for $t_p = 60$ min, and two data fits. The salient point is that no viable difference was detected between the MR traces with DNP for $B_∥ = 0$, $B_∥ = 0.1$ T and $B_∥ = 1.0$ T; the identical data fitting to these three traces bears this out. However, the 3 MR traces with DNP and its fit differ from the MR trace without DNP and its fit, in the expected manner. Hence, application of $B_∥$ does not change the DNP process. The reason lies in the fact that the nuclear spin relaxation due to $B_L$ is already amply suppressed by $B_e \gg B_L$ and hence application of $B_∥$ does not measurably add additional suppression.

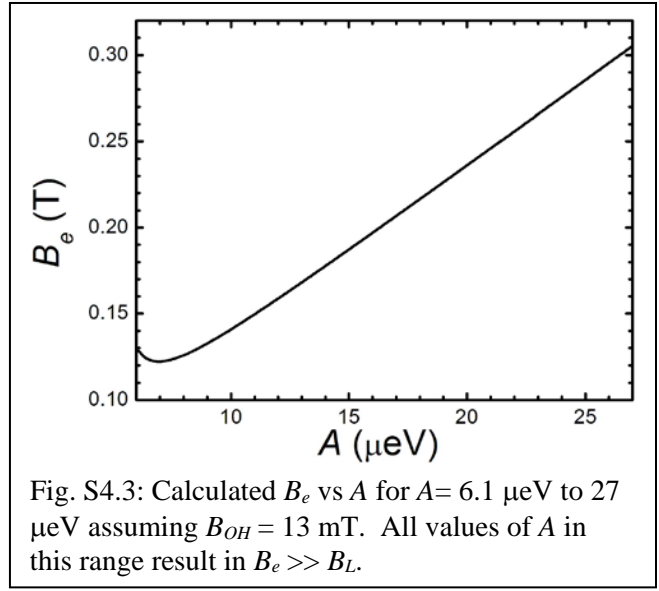

Fig. S4.3: Calculated $B_e$ vs $A$ for $A = 6.1$ µeV to 27 µeV assuming $B_{OH} = 13$ mT. All values of $A$ in this range result in $B_e \gg B_L$.

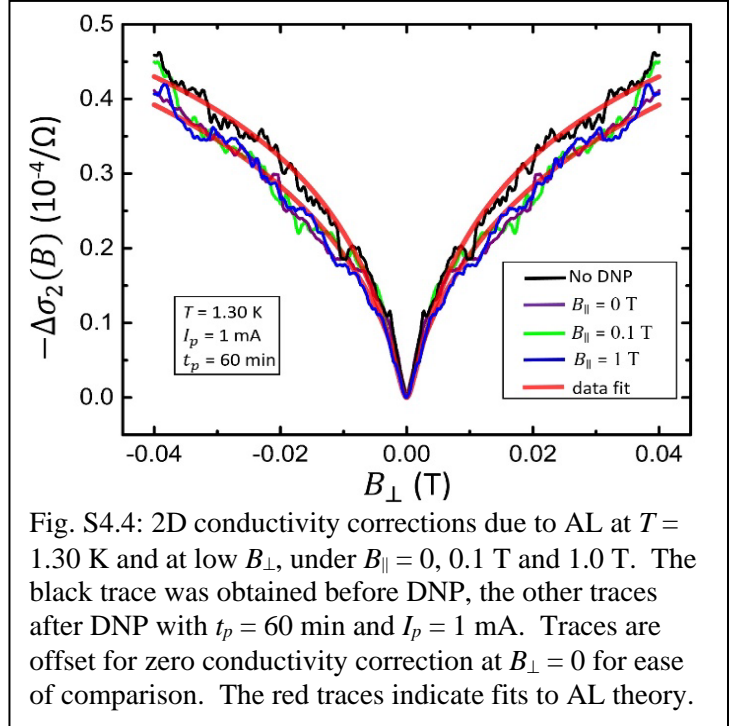

Fig. S4.4: 2D conductivity corrections due to AL at $T = 1.30$ K and at low $B_⊥$, under $B_∥ = 0$, 0.1 T and 1.0 T. The black trace was obtained before DNP, the other traces after DNP with $t_p = 60$ min and $I_p = 1$ mA. Traces are offset for zero conductivity correction at $B_⊥ = 0$ for ease of comparison. The red traces indicate fits to AL theory.

The data in Fig. S4.4 was obtained at $T = 1.30$ K to accommodate the in-plane magnetic fields. The AL data fit under DNP results in: $\tau_\phi = 0.118$ ns, $\tau_{SO} = 1.35$ ps, $B_{OH} = 12.65$ mT. We note that the $B_{OH}$ at $T = 1.30$ K is lower than its saturation value of 13 mT at 0.39 K (main text). We attribute this to the influence of thermal broadening of the AL MR at the higher $T$, expressed in a lowering of $\tau_\phi$ for rising $T$ as expected. It should not be interpreted as a drop in $B_{OH}$ with rising $T$. The characterization of DNP by the AL method requires low $T$ so that the thermal lowering of $\tau_\phi$ does not obscure the lowering of $\tau_\phi$ due to $B_{OH}$.

## S5: Dependence on polarization current, and conversion efficiency

The in-plane Overhauser field $B_{OH}$ shows a dependence on the DC polarization current $I_p$, as we show in this section. DNP experiments were performed with $I_p$ = 0.5 mA, 1 mA, and 1.5 mA at fixed polarization duration $t_p$ = 60 min, at $T$ = 0.39 K. $I_p \geq 2$ mA was not used, because with $I_p$ = 2 mA a momentary small rise of $T$ by about 10 mK was observed for about 15 s after applying $I_p$, and hence it could not be fully ascertained that sample heating was negligible. Figure S5.1 depicts AL MR (negative of conductivity correction $\Delta\sigma_2$) plotted as $-\Delta\sigma_2(B_\perp)$ vs $B_\perp$, where $B_\perp$ denotes the magnetic applied normally to the surface, parametrized in $I_p$ = 0 mA (no DNP), 0.5 mA, 1 mA, and 1.5 mA ($t_p$ = 60 min). The data at $I_p$ = 0 mA denotes a measurement without DNP. Best fits to the ILP theory [10] are indicated as red lines and allow reliable extraction of values for $\tau_{SO}$ and $\tau_\phi$.

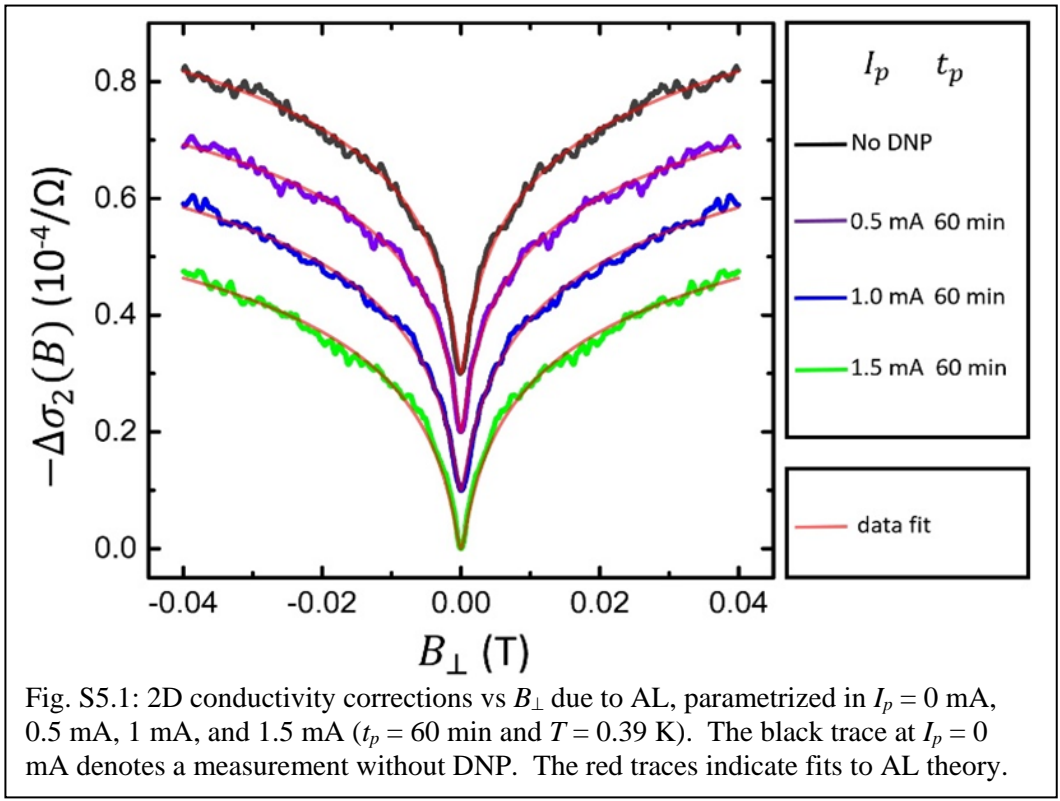

Fig. S5.1: 2D conductivity corrections vs $B_\perp$ due to AL, parametrized in $I_p$ = 0 mA, 0.5 mA, 1 mA, and 1.5 mA ($t_p$ = 60 min and $T$ = 0.39 K). The black trace at $I_p$ = 0 mA denotes a measurement without DNP. The red traces indicate fits to AL theory.

The dependences of $\tau_{SO}$ and $\tau_\phi$ on $I_p$ at $T$ = 0.39 K are presented in Fig. S5.2. The value of $\tau_{SO}$ increases sublinearly with increasing $I_p$, while the value of $\tau_\phi$ decreases with increasing $I_p$. Both are due to increased influence of $B_{OH}$ after using higher $I_p$. The decrease in $\tau_\phi$ is directly linked to higher $B_{OH}$ with higher $I_p$.

The estimated average value of $B_{OH}$ was calculated from $\tau_\phi$ using Eq. 1 (main text) and plotted vs $I_p$ in Fig. S5.3. $B_{OH}$ increases sublinearly with $I_p$ but does not saturate for $I_p \leq 1.5$ mA. The results in Fig. S5.3 are consistent with increasing $I_p$ leading to increasing nuclear polarization, as expected. Figure S5.3 shows that $B_{OH}$ vs $I_p$ strongly resembles a Brillouin function shape, which can be understood from the discussion in Supplemental S4.

We denoted $S_{av}$ the average electron spin polarization due to the Edelstein effect (Supplemental S4). $S_{av}$ is due to charge-current to spin conversion via the Edelstein effect and $S_{av}$ is expected to depend linearly on $I_p$ with a proportionality constant describing the charge-current to spin conversion efficiency, $S_{av} = \alpha\, I_p$. Then, since $B_{OH}$ vs $S_{av}$ follows a Brillouin function, it is expected that $B_{OH}$ vs $I_p$ also follows a Brillouin function. However, a qualitative predictive model linking $B_{OH}$ and $I_p$ depends on either knowledge of a specific value for the hyperfine constant $A$ or of the proportionality constant $\alpha$. The latter

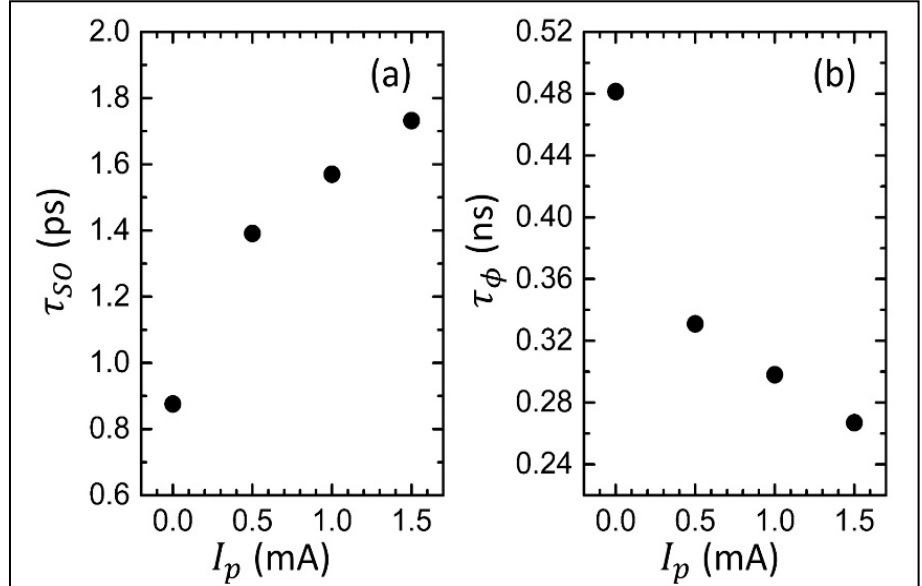

Fig. S5.2: (a) Spin-orbit decoherence times $\tau_{SO}$ and (b) quantum phase decoherence times $\tau_\phi$ ($t_p$ = 60 min and $T$ = 0.39 K), plotted vs different polarization currents $I_p$. The data without DNP is indicated as $I_p = 0$.

will in our samples not only depend on the intrinsic charge-current to spin conversion efficiency of the Edelstein effect, but will also depend on sample geometry due to current spreading. As explained, the calculated $B_{OH}$ reflects a spatial averaging due to current spreading over the sample geometry between the two current contacts, which likely results in non-uniform DNP. The hyperfine constant $A$ is from the literature only known within a range, $A \approx 6.1\ \mu eV\ ...\ 27\ \mu eV$ [16,18-20]. For every given $B_{OH}$ (given $I_p$) we

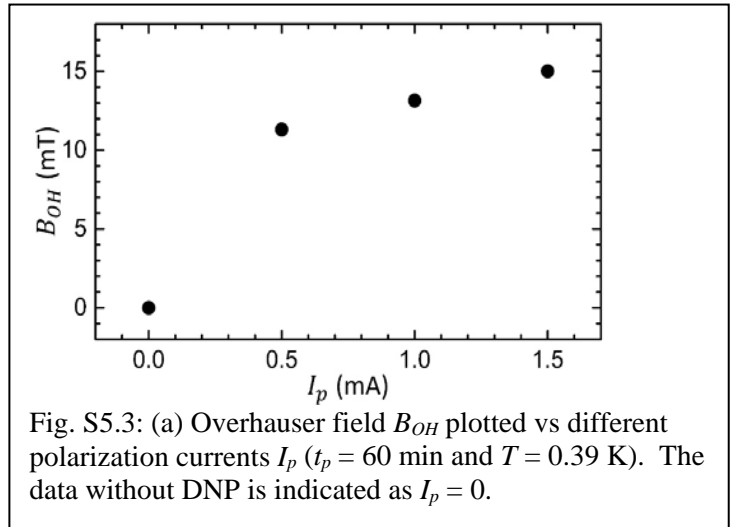

Fig. S5.3: (a) Overhauser field $B_{OH}$ plotted vs different polarization currents $I_p$ ($t_p$ = 60 min and $T$ = 0.39 K). The data without DNP is indicated as $I_p = 0$.

can calculate a range of $S_{av}$ depending on $A$. From this range of $S_{av}$ we can then also only obtain a range of $\alpha$ and not a specific value. Given the uncertainty in $A$ and the spatially non-uniform DNP, a precise analysis of charge-current to spin conversion efficiency and of $\alpha$ would hence be conjectural. Yet, the resemblance to the expected Brillouin function shows a strong consistency between the theoretical expectations and our results.

## S6: Dependence on delay time: magnetoresistance data

Figure S6.1 contains the AL MR as the negative of the 2D conductivity correction $\Delta\sigma_2$ plotted as $-\Delta\sigma_2(B_\perp)$ vs $B_\perp$, with $B_\perp$ the magnetic applied normally to the surface, parametrized in the delay time $t_{delay}$ = 15 min, 20 min, 30 min and 40 min. The data labeled $t_{delay} \to \infty$ denotes a measurement without DNP (returning to a state where nuclear polarization has decayed). The data was obtained at $I_p$ = 1 mA, $t_p$ = 60 min and $T$ = 0.39 K. Best fits to the AL theory [10] are indicated as red lines and allow for determination of values for $\tau_{SO}$ and $\tau_\phi$. The dependences on $t_{delay}$ of $\tau_{SO}$, $\tau_\phi$ and $B_{OH}$ are presented and discussed in the main text.

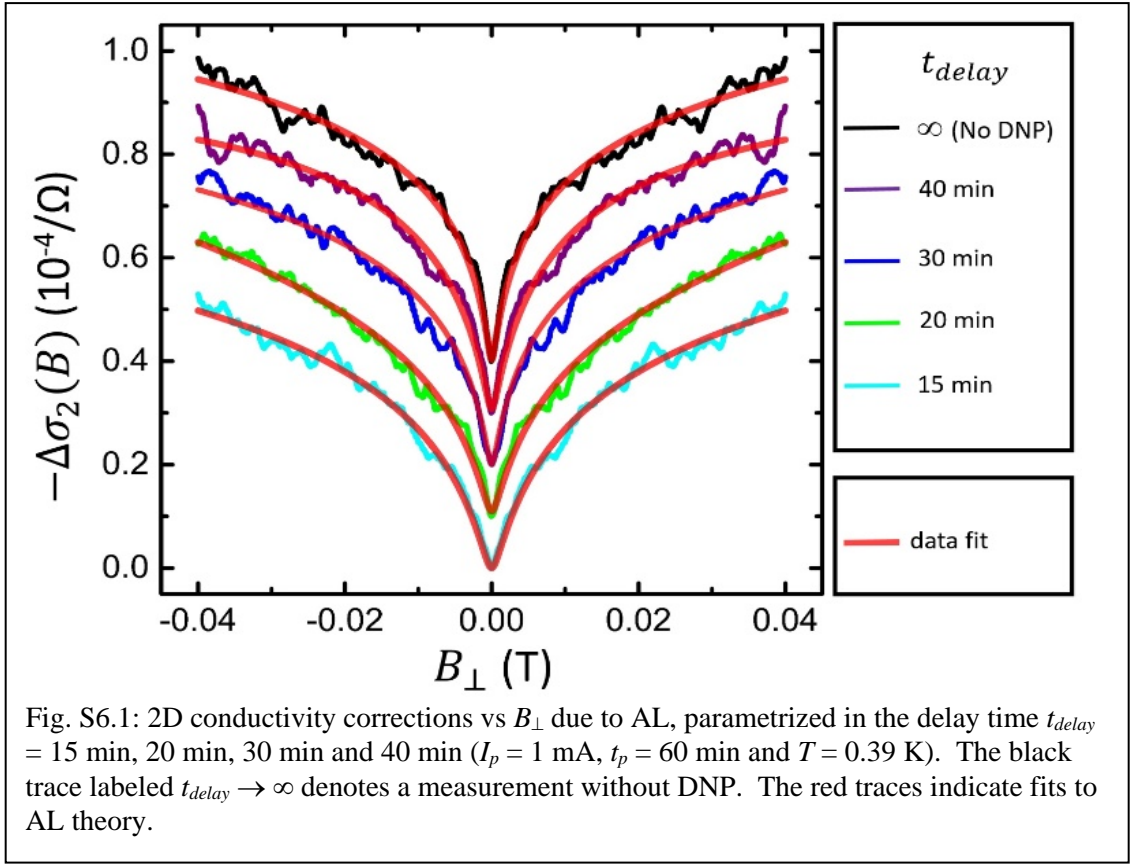

Fig. S6.1: 2D conductivity corrections vs $B_\perp$ due to AL, parametrized in the delay time $t_{delay}$ = 15 min, 20 min, 30 min and 40 min ($I_p$ = 1 mA, $t_p$ = 60 min and $T$ = 0.39 K). The black trace labeled $t_{delay} \to \infty$ denotes a measurement without DNP. The red traces indicate fits to AL theory.


\fi

\end{document}